# Can the electronic energy spectra of *bulk* excitonic states above the lowest energy exciton be traced to parent molecular states in fluorene and its hetero-analogues?

*It depends on the mutual orientation of the corresponding transition dipoles in the crystal.*

Lina Nakhimovsky, (formerly -Visiting Scientist, Chemistry Department, Stony Brook University, Stony Brook, NY 11794), *linanakh@yahoo.com*

***Abstract***

The effect of intermolecular interactions on intensity redistribution between individual electronic transitions in different polymorphs of fluorene and dibenzofuran was studied by *transmittance* electronic spectroscopy in the energy range from onset of absorption to the ionization energy (nine or ten electronic transitions for each polymorph). Electronic transitions with transition dipoles at oblique angles to the crystallographic axes ($\boldsymbol{A_1}$ symmetry) in these crystals resemble molecular spectra. Transitions of $\boldsymbol{B_2}$ symmetry, which in these crystals have parallel transition dipoles, resemble molecular spectra for *surface (substrate influenced)* states. For *bulk* excitonic states the spectra of these transitions have a continuous, uniform intensity distribution in the whole energy range investigated, except when the $\boldsymbol{B_2}$ transition is at the onset of absorption in which case the corresponding spectrum can be traced to the parent molecular state .

***Introduction***

Organic π−conjugated crystals and polymers are extensively studied for their uses in opto-electronic devices and for prospective uses in solar cells. Understanding the electronic properties of these compounds is crucial from a fundamental point of view, and for optimizing the performance of such devices.

It is widely believed that electronic excitations (electron-hole pairs) are localized on a single molecule (Frenkel excitons) or on nearby molecules in the crystal (charge transfer excitons) - models that imply some degree of resemblance between the crystal and corresponding molecular spectra. This perception is based on experimental studies of electronic transmittance spectra of numerous organic crystals, but in most cases only transitions to low energy states were investigated. Also, in the crystals extensively studied (e.g., anthracene, polythiophenes) the optical indicatrix axes do not coincide with the crystallographic axes, thus complicating assignment of the electronic bands by symmetry in the crystal; as a result the comparison with corresponding transitions in the molecular spectra is hindered. Intensity redistribution between multiple electronic transitions, when $\pi$-conjugated molecules join to form a crystal, is the most neglected problem concerning electronic properties of organic solids, the result of which is a dismal lack of information on the subject in textbooks. This work is an attempt to begin filling in the existing knowledge gap.

Numerous theoretical and some experimental studies of intensity redistribution in electronic spectra of molecular associations (such as biopolymers, polymers with aromatic pendants, dimers, and higher aggregates of aromatic molecules) are reported in the literature. It was demonstrated that substantial intensity redistribution between electronic transitions can occur when molecules join to form a molecular aggregate (hypochromism - diminished intensity, and hyperchromism - increased intensity of an individual transition in question) [1–13]

Several different approaches have been used for the theoretical interpretation of hypochromic and hyperchromic effects in molecular associations. The first-order perturbation theory was originally employed by Tinoco [1] and Rhodes [2] to explain the experimentally observed hypochromism in helical DNA. For low-energy electronic transitions this theory predicts hypochromism, if the transition dipoles of the monomers are parallel to each other and perpendicular to the line connecting the centers of monomers (such arrangement will be referred to as parallel arrangement of transition dipoles). In agreement with the sum rule for the oscillator strengths, hyperchromism for higher energy transitions is predicted. In case of collinear arrangement of the monomeric transition dipoles, theory predicts hyperchromism for low energy transitions (and hypochromism at higher energies). Local field theory [3], the time-dependent self-consistent field theory [4], linear response functions method [5] were also employed for predicting the effect of molecular interactions on intensity redistribution between transitions in electronic absorption spectra of molecular aggregates.

Second quantization method developed by Agranovich [6] can be used for calculating the oscillator strengths redistribution in molecular associations. Hoffmann [7] applied the latter theory to calculating the hypo/hyperchromic effect in a one-dimensional crystal with one molecule per unit cell. The predictions of the second quantization method are qualitatively similar to those of the previous theories, but it was shown in Ref. 7 that the magnitude of hypo- or hyperchromism calculated in the approximation of the first-order perturbation theory could differ by approximately 50% from that predicted by the more inclusive second quantization method.

Experimentally, hypo/hyperchromism was detected in biopolymers (DNA and proteins) [1,8], in dimers and other associations of aromatic molecules in amorphous [9] and Shpol'sky matrices [10], in poly(N-vinylcarbazole) [11,12], and in other polymers with aromatic pendants [11]. Hypochromism was also observed in the absorption spectrum of molecules with a carbazole moiety after their incorporation into circular DNA [13].

Fluorene and its heteroanalogues present useful models for comprehensive studies of the effect of sample preparation on the electronic properties of π-conjugated organic crystals, particularly on the intensity redistribution between multiple electronic transitions, when molecules join to form a crystal. The molecular structures of fluorene and its hetero-analogues are very similar, as are their crystalline structures. Transitions of two symmetry types - $\boldsymbol{A_1}$ and $\boldsymbol{B_2}$ - are allowed in their molecular (gas phase, solution) electronic absorption spectra. Assignment of molecular electronic transitions (absorption bands) of these compounds by symmetry in the near ultraviolet energy region is discussed in Refs.14-19. In a more recent paper the assignments of electronic transitions (bands) for fluorene and three of its hetero-analogues were extended to the ionization energy [20]. These assignments were based on experimental studies of transmittance spectra of the gas and crystalline phases and of linear dichroism spectra of the corresponding molecules embedded in stretched polymer sheets. Theoretical calculations of electronic properties of these molecules by the time-dependent density functional theory TD-B3LYP/6-31 +G(d,p) are in agreement with the general trends in the experimental spectra but the numerical agreement between experimental and theoretical data is not always satisfactory [20]. Electron energy loss (EEL) spectra of fluorene vapor and single crystals were measured in work [21]. Detailed comparison between our data and those of Ref. 21 is not possible because of low resolution of the EEL spectra, but the latter corroborate our results that at energies between 7 and 8 eV transitions of $\boldsymbol{A_1}$ symmetry dominate. The electronic properties of crystalline fluorene as a function of pressure were studied theoretically in Ref. 29. As the authors of work [29] remark, their results are not in agreement with experimental spectra but their interest was in the effect of pressure on the spectra. Polarized electronic spectra of fuorene and dibenzofuran crystals in the near UV region were reported in Ref. 14, 15 and 28. Our assignments of electronic transitions of $\boldsymbol{A_1}$ symmetry in the near UV region agree with those in Ref 14 and 15.

*Transmittance* spectra of these compounds in the c crystallographic direction (transitions of $\boldsymbol{B_2}$ symmetry) were not reported in the literature previously, their reflection spectra were studied in Ref. 18.

Several compounds in this group (i.e., fluorene, carbasole and dibenzofuran) form orthorhombic crystals with four molecules per unit cell [22-24], space group $P_{nma}$ (Figure 1, panel I). The transition dipoles of the $\boldsymbol{A_1}$ symmetry transitions ($B_{2u}$ and $B_{3u}$ in the crystal) lie in the ab plane of the crystal at oblique angles to the a and b crystallographic axes; those of $\boldsymbol{B_2}$ symmetry ($B_{1u}$ in the crystal) are perpendicular to the ab plane of the crystal and parallel to each other and to the c crystallographic axis. This difference in mutual orientation of electronic transition dipoles in the crystals studied will be helpful in discerning its effect on the character of the respective electronic absorption spectra.

Molecules in these crystals form parallel layers within which the closest distance between molecules is around 4Å, while the closest distance between the centers of molecules in adjacent layers is around 10Å (Figure 1, panel I). Thus interaction between the molecules within the layers is much stronger than between layers; therefore, to a first approximation, these crystals can be considered quasi-two-dimensional.

Thin (0.1-0.3 μ thickness) crystals (sublimation flakes) of these compounds grow in the ab plane, so that at normal incidence of light on the sample only $\boldsymbol{A_1}$ ($B_{2u}$, $B_{3u}$) symmetry transitions can be observed in their transmittance spectra [14-16, 25], allowing unambiguous assignments of transitions of this symmetry in the crystal. At normal incidence of light transitions of $\boldsymbol{B_2}$ symmetry cannot be detected in crystals with a developed ab plane (Fig. 1, panel II).

In this paper we present electronic transmittance spectra of two members of the group - fluorene and dibenzofuran - in the energy range of nine to ten electronic transitions for each of several polymorphs of these compounds

***Experimental***

Fluorene (Supelco #48568) and dibenzofuran (OEKANAL, analytical standard) were obtained from commercial sources. We have measured the electronic spectra of six different types of samples of fluorene, which presented at least four modifications (polymorphs) with distinct transmittance spectra. Type 1 was obtained by melting the material between LiF plates and applying strong uniaxial pressure in the direction perpendicular to the surface of the sample during the crystallization of the melt. These samples grow in the ab plane with thicknesses less than 0.1 μ. Sample 2 was a sublimation flake of thickness around 0.2 μ, obtained by heating the material in a sealed Petri dish in the presence $N_2$ gas; this sample also had a developed ab plane. As discussed in Ref. 15, thin sublimation flakes develop pinholes in open air. To prevent the formation of pinholes the flakes were hermetically sealed between LiF plates using Duco cement. Samples 3 and 4 were sublimation flakes of smoky color (which implies large birefringence) with thickness of around 0.90 μ. Type 5 was obtained from melt with moderate uniaxial pressure, in which case the estimated thickness of samples was around 0.25 μ. Type 6 was obtained from melt under fast crystallization conditions (thickness about 0.40 μ).

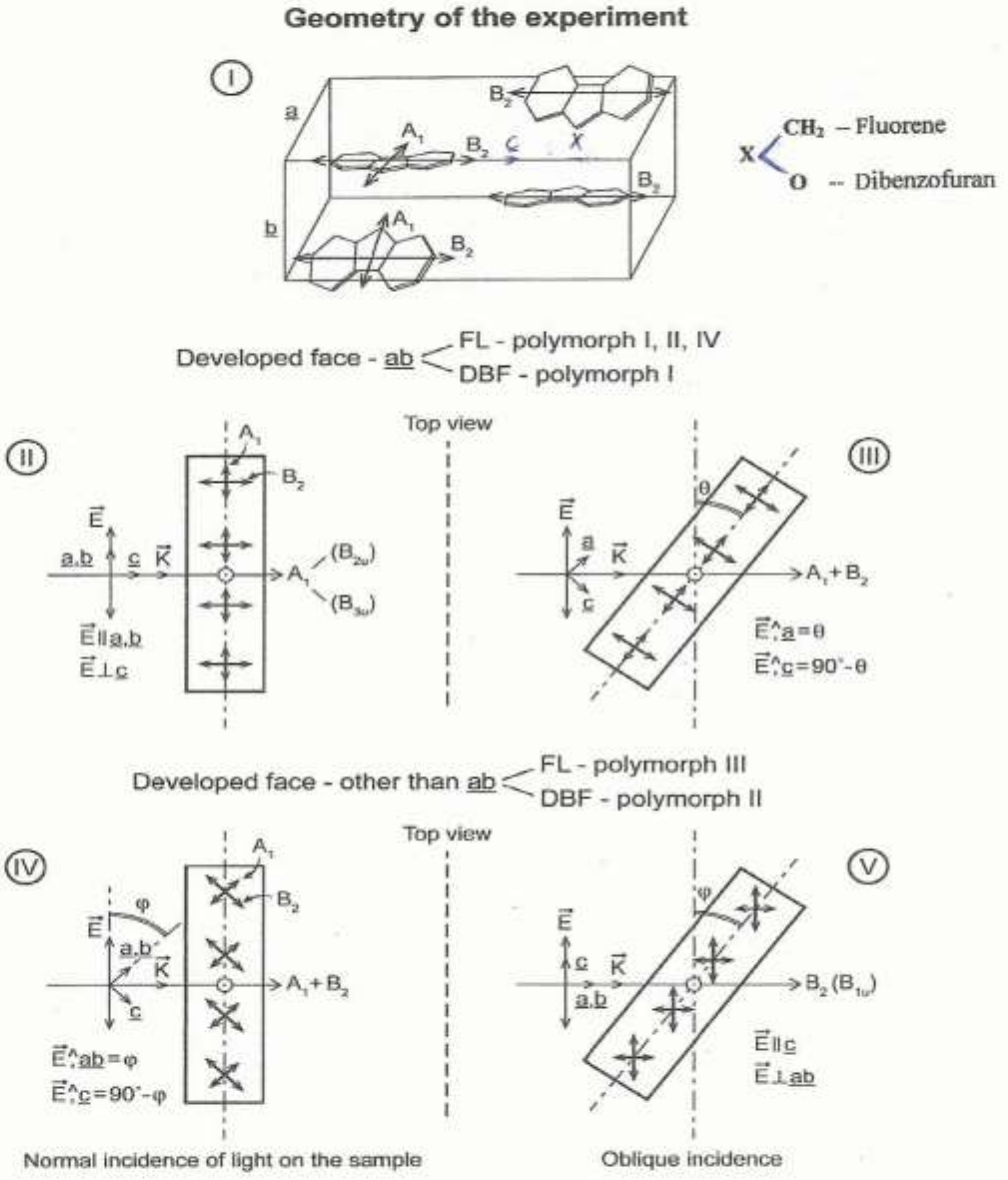

**FIGURE 1.**
Geometry of the experiment: I- crystalline structure of fluorene and dibenzofuran; II- normal incidence of light on the ab plane of the sample; III- oblique incidence of light on the ab plane of the sample; IV- normal incidence of light on a developed plane containing a projection of the crystallographic direction; V- oblique incidence of light on a developed plane containing a projection of the c crystallographic direction.

Transmittance spectra of three types of samples of dibenzofuran were also measured. Type 1 was grown under strong uniaxial pressure, type 2 was a sublimation flake. The thicknesses of these samples varied between 0.20 and 0.30 μ. Sample type 3 was grown from melt under fast crystallization conditions (thickness ~ 0.5 μ)

The spectra were recorded at the Synchrotron Light Source at Brookhaven National Laboratory. A modified Wadsworth monochromator scans through the wavelength region of interest. The photomultiplier tube has a magnesium fluoride window transparent to about 120 nm; a transimpedance amplifier (ORIEL) converts the PMT signal to voltage. The computer interacts with a lock-in amplifier, the samples, the high voltage and synchrotron ring signals. Neglecting reflectance, the data for the sample and the respective baseline allow the calculation of *Absorbance (log 1/T)* of the crystal as a function of wavelength.

The thicknesses of samples were estimated from measurements of birefringence, using a polarizing microscope fitted with a U-CSE Olympus compensator, an atomic force microscope, and using spectroscopic data from Refs. 14 and 15. The discrepancy in thickness values between different methods was about 30%, thus the absolute intensities measurements are only semi-quantitative.

Spectra of sample 6 of fluorene were measured in $N_2$ gas, and of sample 3 of dibenzofuran in argon gas, while spectra of all other samples were measured in vacuum.

To establish if the observed spectra have molecular character the crystal spectra were compared with the corresponding synchrotron radiation linear dichroism (SRLD) spectra from Ref. 20.

## ***Results and discussion***

*Fluorene.* In Figure 2 the electronic transmittance spectra of samples 1 and 2 are shown. The geometry of the experiment is shown in Fig. 1, panels II and III. The two factor-group (Davidov) components [27] of the $B_{2u}$ and $B_{3u}$ ($\boldsymbol{A_1}$) symmetry electronic transitions of sample 1 (polymorph I) are depicted in Figure 2, spectra I and II. As suggested in Ref. 20 the two closely spaced bands near 210 nm (5.95 eV) in the synchrotron radiation linear dichroism (SRLD) spectrum (Fig. 2, spectrum Ia; reproduced from Ref. 20) correspond to 3$\boldsymbol{A_1}$ and 4$\boldsymbol{A_1}$ transitions. Similarly in polymorph I they form a single band, and the spectrum of the **E**II<u>a</u> component is almost identical with the spectrum of the $\boldsymbol{A_1}$ symmetry transitions of the SRLD (molecular) spectrum of fluorene. Thus in the spectrum of polymorph I of fluorene at least three lowest energy transitions of this symmetry closely resemble molecular spectra (predominantly Frenkel excitons). The factor-group splitting of all states is within experimental error with the exception of the 6$\boldsymbol{A_1}$ transition, for which the exact maximum position of the **E**ll<u>a</u> component could not be determined, but it is at substantially higher energy than that of the **E**ll<u>b</u> component. It is interesting, that the <u>a</u>-component of the 2$\boldsymbol{A_1}$ transition is extremely weak, similar its intensity in the SRLD spectrum while in the <u>b</u>-component is almost an order of magnitude stronger, similar to the intensity of this transition in the polymorph II of fluorene (spectra V and VI of Fig. 2). A possible explanation of this observation could be that the sample corresponding to polymorph I was

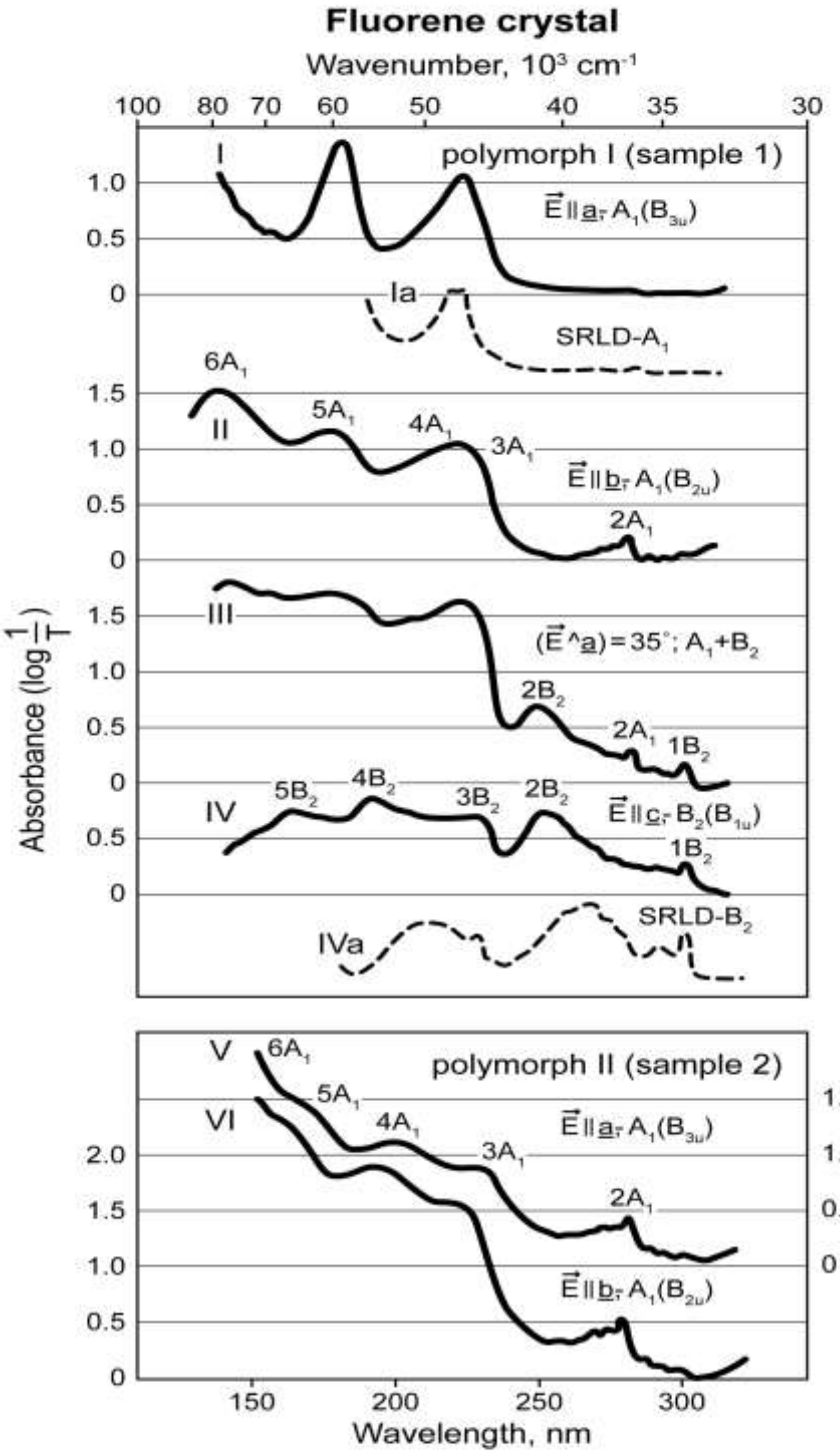

produced

**FIGURE 2.** Electronic transmittance spectra of samples 1 and 2 of fluorene. Sample 1 (*polymorph I*):I- **E**ll<u>a</u>; Ia- SRLD (linear dichroism) $A_1$ symmetry spectrum of fluorene; II- **E**ll<u>b</u>; III- (**E**^<u>a</u>)= $35^0$ ; IV- **E**ll<u>c</u>; IVa- $B_2$ symmetry linear dichroism spectrum of fluorene. Sample 2 (*polymorph II*): V- **E**ll<u>a</u>; VI- **E**ll<u>b.</u>

produced under uniaxial pressure which could have led to increase in the intermolecular distance in the a crystallographic direction.

As expected, long axis polarized transitions are barely detectable at normal incidence of light on sample 1 of fluorene. In order to obtain a crystal spectrum containing transitions of $\boldsymbol{B_2}$ ($B_{1u}$) symmetry (along with the $\boldsymbol{A_1}$ symmetry transitions) sample 1 was rotated $35^0$ around the axis perpendicular to the direction of the electric field vector **E** (Fig. 1, panel III), and the transmittance spectrum of the sample was measured (at oblique incidence of light on the sample surface, Fig. 2, spectrum III). The $1\boldsymbol{B_2}$ and $2\boldsymbol{B_2}$ symmetry transitions - at around 300 and 245 nm (4.17 and 5.10 eV) - are clearly seen in the spectrum of the tilted crystal. The higher energy $\boldsymbol{B_2}$ symmetry transitions are overlapped by $\boldsymbol{A_1}$ symmetry transitions, and the maxima of the former are not readily identifiable in this spectrum. In order to separate the transitions of different symmetries, the b-component of the spectrum of the crystal at normal incidence of light (which consists of only $\boldsymbol{A_1}$ symmetry transitions) was subtracted from the spectrum of the tilted crystal (which consists of transitions of both symmetries). The spectrum of the b-component was so scaled that the intensity of the $2\boldsymbol{A_1}$ transition pure electronic band (at 280 nm, 4.46 eV) becomes zero in the subtracted spectrum (Fig. 2, spectrum IV). Comparison of the latter with the SRLD spectrum of the $\boldsymbol{B_2}$ symmetry transitions (Fig. 2, spectrum IVa) shows that for each of the four electronic transitions of $\boldsymbol{B_2}$ symmetry in the SRLD (molecular) spectrum there is a corresponding transition in the spectrum of polymorph I of fluorene. Consequently the excitations to all electronic states of both symmetries in the spectrum of polymorph I can be interpreted as predominantly due to Frenkel excitons and possibly as a combination of Frenkel and charge transfer excitons.

Similar to polymorph I, five $\boldsymbol{A_1}$ symmetry electronic transitions are detected for a thin *sublimation* flake (Fig. 2, compare spectra I and V), but the positions of maxima and intensity distribution between transitions in samples 1 and 2 are substantially different. Therefore, we assume that sample 2 presents a new polymorph of fluorene (polymorph II). The factor-group splitting for all ***$A_1$*** symmetry transitions is again within experimental error. The positions of maxima and the intensities of ***$A_1$*** transitions for polymorphs I and II are shown in Table 1. (In parenthesis the maximum value of Absorbance (log 1/T) of each electronic band is shown)

| Electronic transition | Fluorene | | | | | | Dibenzofuran | | | | | |
|---|---|---|---|---|---|---|---|---|---|---|---|---|
| Symmetry | Polymorph I | | | Polymorph II | | | Polymorph I | | | Polymorph II | | |
| | nm | ***eV*** | *A* | nm | ***eV*** | *A* | nm | ***eV*** | *A* | nm | ***eV*** | *A* |
| ***$2A_1$*** $(B_{,3u})$ | 282 | ***4.43*** | (0.0 1) | 282 | ***4.43*** | (*0.25*) | 306 | ***4.10*** | (*0.63*) | 306 | ***4.10*** | (*0.43*) |
| $(B_{2u})$ | 282 | | (*0.08*) | *282* | | *(0.45)* | 306 | | (*1.29*) | 306 | | (*0.94*) |
| ***$3A_1$*** | | | | 229 | ***5.46*** | (0.55) | 229 | ***5.46*** | (*0.98*) | 229 | ***5.46*** | (*0.56*) |
| $(B_{,3u})$ | 218 | ***5 .73*** | (*0.50*) | 229 | | (*1.00*) | 229 | | (*1.86*) | 229 | | (*1.26*) |
| $(B_{2u})$ | 218 | | (*0.45*) | | | | | | | | | |
| ***$4A_1$*** | | | | 195 | ***6.41*** | (*0.85*) | 205 | ***6.09*** | (0.90) | 205 | ***6.09*** | (*0.70*) |
| | | | | 195 | | *(1.90)* | 205 | | (*1.85*) | 205 | | (*1.16*) |
| ***$5A_1$*** $(B_{3u})$ | 172 | ***7.26*** | (*0.65*) | 166 | ***7.53*** | (*1.35*) | 195 | ***6.41*** | (*1.53*) | 165 | ***7.58*** | (*1.57)* |
| $(B_{2u})$ | 172 | | (*0.49*) | 166 | | (*1.90*) | 195 | | (*1.81*) | 165 | | (*2.31*) |
| ***$6A_1$*** $(B_{,3u})$ | > 130 | | (?) | <150 | | (>*1.70*) | 151 | ***8.28*** | (*1.51*) | >130 | | (>*1.87*) |
| $(B_{2u})$ | 129 | ***9.68*** | (*0.72*) | | ***8.33*** | (>*2.30*) | | | (*2.31*) | | ***9.62*** | (>*2.10*) |
| ***$7A_1$*** $(B_{,3u})$ | | | | | | | >126 | | (>*1.69*) | | | |
| $(B_{2u})$ | | | | | | | | ***9.92*** | (>2.54) | | | |
| | nm | ***eV*** | *A* | | | | nm | ***eV*** | *A* | | | |
| ***$1B_2$*** $(B_{1u})$ | 303 | ***4.13*** | (*0.24*) | | | | 273 | ***4.57*** | (*0.29*) | | | |
| ***$2B_2$*** $(B_{1u})$ | 248 | ***5.04*** | (*0.70*) | | | | 252 | ***4.96*** | (*0.24*) | | | |
| ***$3B_2$*** $(B_{1u})$ | 225 | ***5.56*** | (*0.65*) | | | | 214 | ***5.84*** | (*0.63*) | | | |
| ***$4B_2$*** $(B_{1u})$ | 185 | ***6.76*** | (0.*80*) | | | | 190 | ***6.57*** | (*0.94*) | | | |
| ***$5B_2$*** $(B_{1u})$ | 157 | ***7.96*** | (0.69) | | | | 140 | ***8.98*** | (*0.60*) | | | |

We were unable to obtain the spectrum of the $\boldsymbol{B_2}$ ($B_{1u}$) symmetry transitions for polymorph II in this round of experiments.

The large birefringence of sample 3, smoky-colored sublimation flake (~ 1.0 μ thick), led us to assume that the developed plane of this sample contained a large projection of the c crystallographic direction in which case the geometry of the experiment at normal incidence of light on the sample is that of Fig. 1, panel IV. Alignment of the c crystallographic axis with the electric field vector **E** was attempted by rotating the sample around the vertical axis (Fig. 1, panel V). The corresponding spectra are shown in Figure 3. Spectrum I in Fig. 3 was obtained at normal incidence of light with the electric field vector **E** parallel to one of the extinction directions in the crystal. Only the 2$\boldsymbol{A_1}$ transition is recorded at this geometry of the experiment; in the energy range of all other transitions there is complete absorption of light. Rotating the sample $15^0$ around the vertical axis leads to diminished intensity of the 2$\boldsymbol{A_1}$ transition and allows to record absorbance in the energy range of 3$\boldsymbol{A_1}$ and 4$\boldsymbol{A_1}$ transitions (spectrum II in Fig. 3). These two transitions seem to fuse together and form an energy range of uniform intensity, although the possibility of saturation in this energy region can not be excluded (log 1/T= 2.5). Thus the character of the $\boldsymbol{A_1}$ symmetry spectrum in sample 3 could not be determined unambiguously.

Rotating the sample $30^0$ produced the spectrum III of Fig. 3. The intensity of the 2$\boldsymbol{A_1}$ transition dramatically decreased, while the intensity of the transition 1$\boldsymbol{B_2}$ at the onset of absorption substantially increased. All other $\boldsymbol{A_1}$ and $\boldsymbol{B_2}$ symmetry transitions fused together with no clear maxima of transitions of either symmetry up to around 8 eV. Rotating the sample $40^0$ positions the ab crystallographic plane almost perpendicular to the electric field vector (Fig. 1, panel V), so that only a minute intensity of the 2$\boldsymbol{A_1}$ symmetry transition is present. Thus, spectrum IV in Fig. 3 consists of predominantly of $\boldsymbol{B_2}$ ($B_{1u}$) symmetry transitions, and no bands resembling molecular $\boldsymbol{B_2}$ transitions, or the corresponding $B_{1u}$ transitions of polymorph I can be detected. We assume that sample 3 is another polymorph of fluorene (polymorph III).

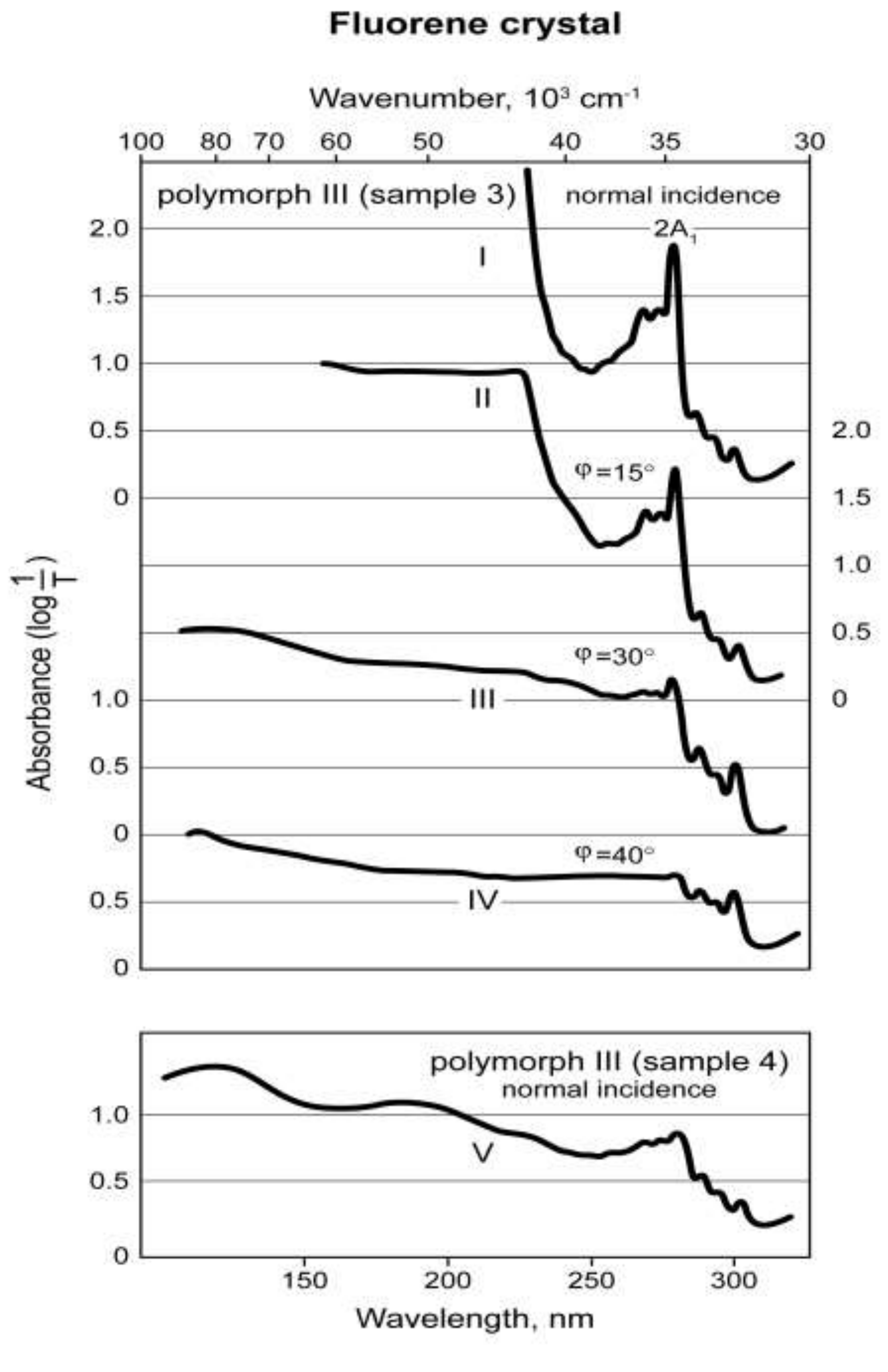

**FIGURE 3**. Electronic transmittance spectra of samples 3 and 4 of fuorene.
Sample 3 ( *polymorph III* ): I- normal incidence of light on the sample;
II- oblique incidence at $15^0$; III- at $30^0$; IV- at $40^0$. Sample 4 (*polymorph III*):
V- spectrum of sample 4 obtained at normal incident of light (see text)

Sample 4 (Fig. 3, spectrum V) is also a smoky-colored sublimation flake, and at *normal incidence* of light has a spectrum similar to that of sample 3 at $30^0$ (Fig. 3, spectrum III) or sample 6 at $55^0$ (Fig 4. spectrum VII), which illustrates that these unusual spectra are not an artifact due to oblique incidence of light on the sample. We assume that this sample is the same polymorph as sample 3 (polymorph III), but it had a developed crystallographic plane with a larger than sample 3 projection of the c crystallographic direction on the electric field vector **E**.

Sample 5, crystallized from melt, grows in a plane close to the ab crystallographic plane and exhibits electronic spectra entirely different from polymorphs I, II and III. It has an $\boldsymbol{A_1}$ symmetry spectrum which in the region 3$\boldsymbol{A_1}$ and 4$\boldsymbol{A_1}$ transitions has a uniform intensity distribution (Figs. 4.I and 4.II). Again, to determine the character of the $\boldsymbol{B_2}$ ($B_{1u}$) symmetry transitions we rotated the sample $50^0$ around the vertical axis. The corresponding spectrum is shown in Fig. 4, spectrum III; subtracting from it the properly scaled **E**llb spectrum produces the spectrum IV in Fig. 4. This spectrum consists of individual electronic transitions of $\boldsymbol{B_2}$ ($B_{1u}$) symmetry, but is distinct from the corresponding spectrum of polymorph I (compare spectrum IV in Fig. 2 with spectrum IV in Fig 4). Thus we assume that sample 5 presents yet another polymorph of fluorene – polymorph IV. It is interesting that the 1$\boldsymbol{B_2}$ and 2$\boldsymbol{B_2}$ transitions in polymorph IV are similar in position and relative intensity to the corresponding transitions in the absorption spectra deduced from reflectance spectra [18], while in polymorph I these transitions resemble corresponding transitions in the gas phase spectrum (Ref. 20).

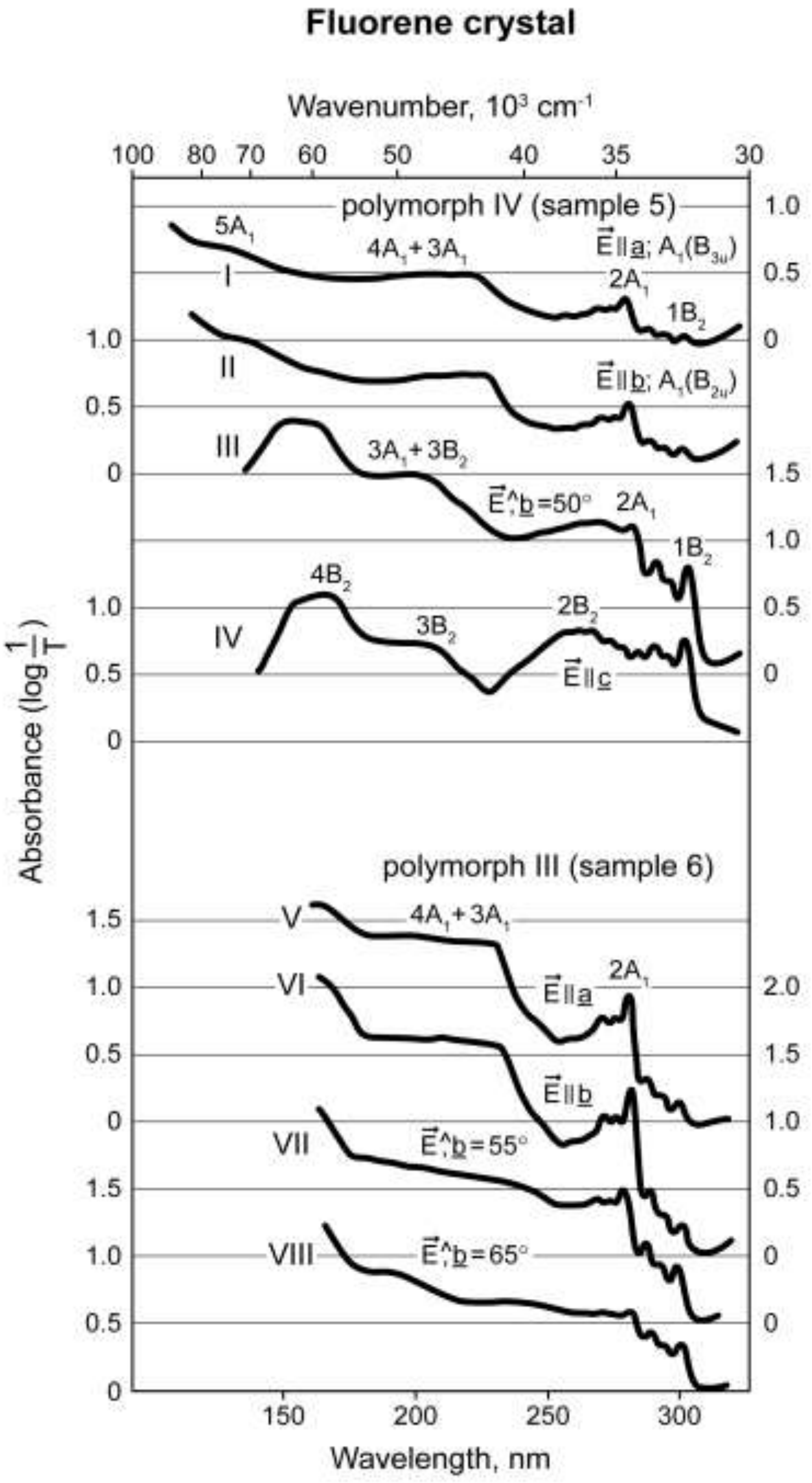

**FIGURE 4**. Electronic transmittance spectra of samples 5 and 6 of fluorene.
*Sample 5 (polymorph IV):* I- **E**lla; II- **E**llb; III- **E**^b = $50^0$; IV- **E**llc (see text).
*Sample 6 (polymorph III)*: V- **E**lla: VI- **E**llb; VII- **E**^b = $55^0$; VIII- **E**^b = $65^0$

Sample 6 is a crystal approximately 0.5 μ thick, obtained under fast crystallization conditions. It produces spectra similar to those of polymorph III (Fig. 4, spectra V, VI, VII and VIII). The spectra of sample 6 were measured in a chamber purged with $N_2$ gas, therefore the spectra are shown to about 165 nm.

Dibenzofuran. At normal incidence of light sample 1 (polymorph I), grown from melt under strong uniaxial pressure, exhibits $\boldsymbol{A}_1$ ($B_{2u}$ and $B_{3u}$) symmetry spectra (only the **E**ll<u>a</u> component is shown; Fig. 5, spectrum II). Comparison of the $\boldsymbol{A_1}$ symmetry spectrum in the crystal with the corresponding SRLD spectrum (Fig 5, panel I, dashed line) shows that, the excitations to 2$\boldsymbol{A_1}$, 3$\boldsymbol{A_1}$, 4$\boldsymbol{A_1}$ and 5$\boldsymbol{A_1}$ states can be interpreted as predominantly due to Frenkel excitons. In Ref. 26 it was shown that the lowest (2$\boldsymbol{A_1}$) transition of the dibenzofuran molecule borrows a substantial part of if its intensity from the overlapping 1$\boldsymbol{B_2}$ symmetry transition via non-totally symmetric vibrations with frequencies around 400 and 700 $cm^{-1}$. As a result, the non-totally symmetric *vibronic* bands in molecular spectra have intensities equal to or larger than the pure electronic band (see *Inset* on top of Fig. 5). When a dibenzofran crystal grows in the <u>ab</u> plane, non-totally symmetric *vibronic* bands cannot be observed at normal incidence of light on the sample (Fig 1, panel II). Since the intensity of the $\boldsymbol{a_1}$ symmetry *vibronic* bands is on the average three times smaller than that of the pure electronic band (see Inset), the ratio of the average intensity in the vibronic band area (290-300 nm) to that in the pure electronic band (306 nm, 4.1 eV) at normal incidence of light on the crystal is expected to be ~1:3, which is the case for sample 1 of dibenzofuran (Fig. 5, panel II)

Rotating the sample by $35^0$ around the axis perpendicular to the electric field vector produced a spectrum of $\boldsymbol{A_1}$ transitions with diminished intensity. The ratio of average intensity of *vibronic* bands to the pure electronic band of the 2$\boldsymbol{A}_1$ transition equals ~1:2. This increase in relative intensity in the area of *vibronic* bands takes place because strong $\boldsymbol{b_2}$ symmetry *vibronic* bands are present in the spectrum of the 2$A_1$ transition. An increase of intensity in the energy ranges corresponding to $\boldsymbol{B_2}$ ($B_{1u}$) symmetry (270-290 nm) is also observed. Employing a procedure similar to that for the polymorph I of fluorene, we obtained a spectrum of four $\boldsymbol{B_2}$ ($B_{1u}$) symmetry transitions of dibenzofuran which exhibit "molecular" character in the whole energy range studied (Fig. 5, panels IV and IVa). Comparison of the *intensities* of these $\boldsymbol{B_2}$ ($B_{1u}$) transitions with the intensities of corresponding transitions in the SRLD spectrum shows that the intensity distribution in the spectrum of the polymorph I of dibenzofuran is different from that in the SRLD spectrum. Unlike the SRLD spectrum, in which all four transitions of $\boldsymbol{B_2}$ symmetry have comparable intensities, in the crystal spectra the lowest energy transition is substantially weaker than those in the high energy region. Thus, hypochromism (diminished intensity) is observed in the low energy transition, and hyperchromism (increased intensity) - in the high energy transitions, in qualitative agreements with theory [1-7, 25].

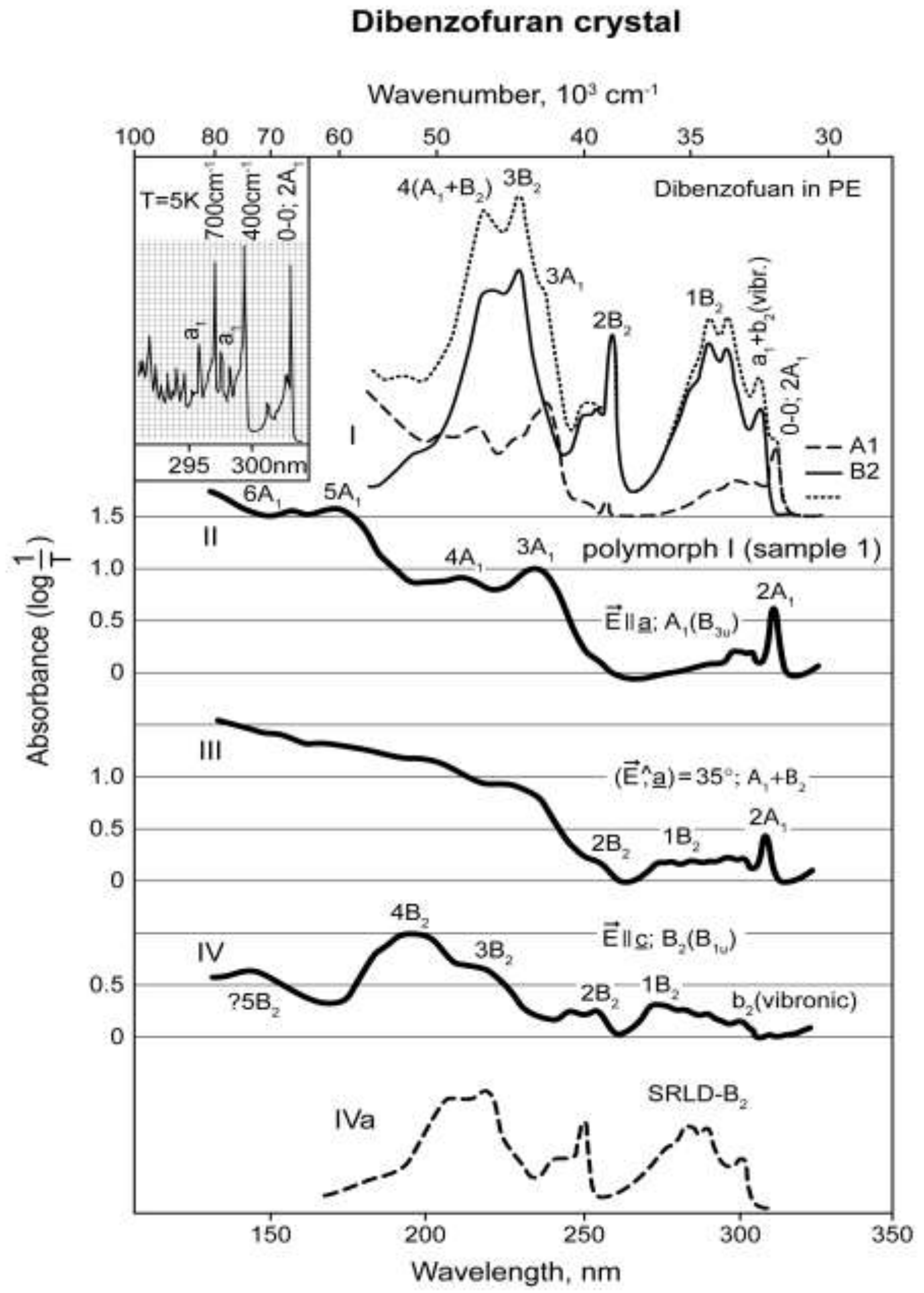

**FIGURE 5**. Electronic transmittance spectra of dibenzofuran.
I- SRLD (molecular) spectrum: ---$A_1$ symmetry, $B_2$ symmetry,
….. combined spectrum. *Insert*- spectrum of the onset of 2$A_1$ transition in fast frozen heptane at 5K.
Electronic transmittance spectra of dibenzofuran crystal. Sample 1 (*polymorph I*): II- **E**ll<u>a</u>; III- **E**^<u>a</u> = 35$^0$; IV- **E**ll<u>c</u> (see text);
IVa- SRLD $B_2$ symmetry spectrum.

At normal incidence of light on a s*ublimation* flake of dibenzofuran (sample 2) the spectra reveal five $\boldsymbol{A_1}$ electronic transitions (Fig. 6, spectra I and II) which are similar to those of polymorph I in the near UV region and substantially different in the high energy region (compare Fig. 5, spectrum II with Fig. 6, spectrum I). We assume that sample 2 presents a new polymorph of dibenzofuran – polymorph II. Comparison of the spectrum of $\boldsymbol{A_1}$ symmetry with the SRLD spectrum (Fig 5, panel I, dashed line) shows that the excitations to 2$\boldsymbol{A_1}$, 3$\boldsymbol{A_1}$, 4$\boldsymbol{A_1}$ and 5$\boldsymbol{A_1}$ states can be interpreted as predominantly Frenkel excitons. The factor-group splitting of all transitions is within the experimental error.

No spectrum of the $\boldsymbol{B_2}$ symmetry transitions was obtained for polymorph II of dibenzofuran. The wavelengths and intensities in the spectra of polymorphs I and II of dibenzofuranare shown in Table 1.

Fast crystallized melt of dibenzofuran (sample 3) consisted of several monocrystals. Spectra of two monocrystals were measured. At normal incidence of light, the spectrum of monocrystal 1

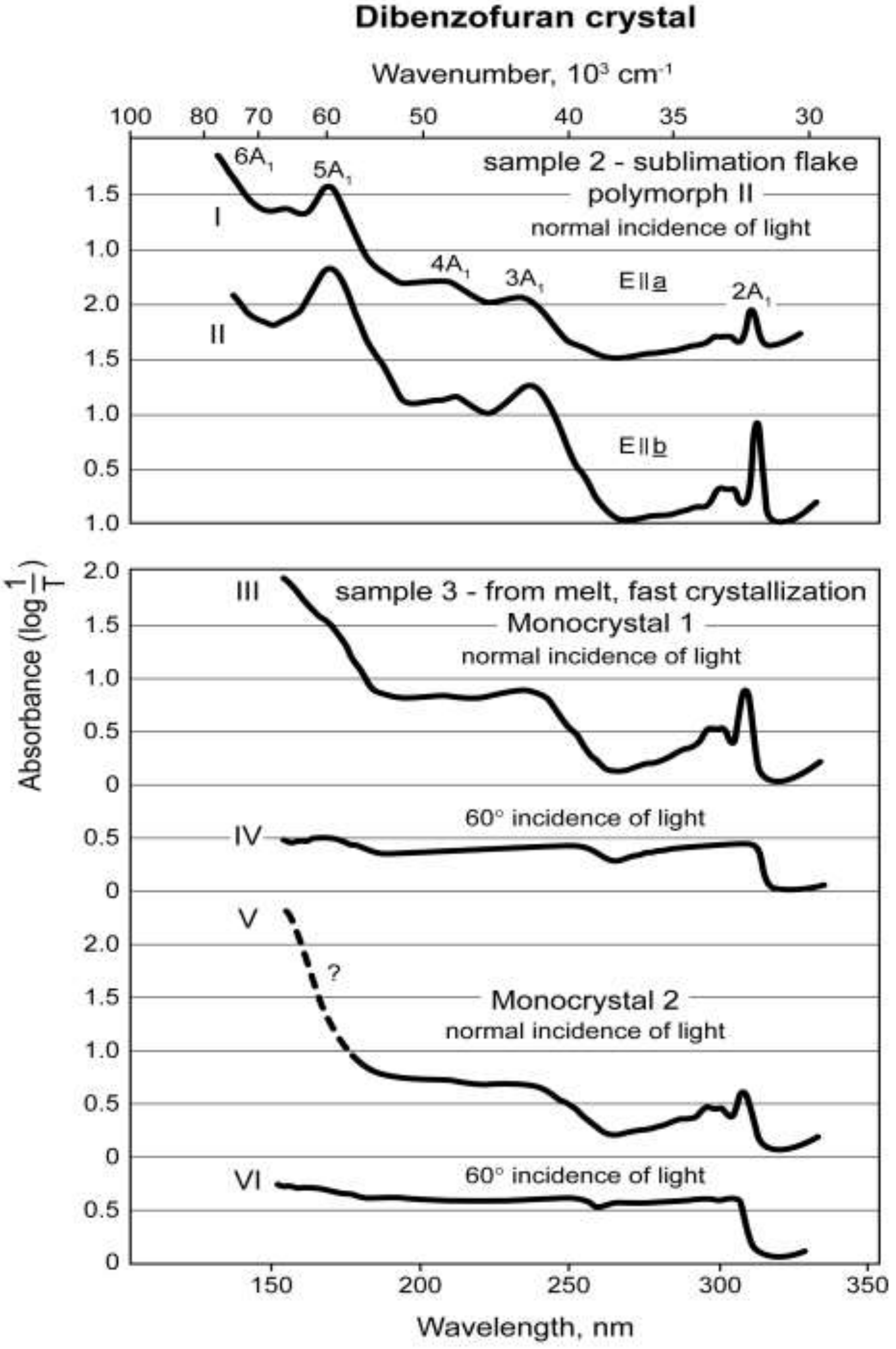

*FIGURE 6.* Electronic spectra of dibenzofuran.
Sample 2 (sublimation flake, *polymorph 2*): I- **E**ll<u>a</u>; II- **E**ll<u>b</u>.
Sample 3 (fast crystallized from melt) --- *monocrystal I:* III- normal incidence of light; IV - $60^0$ incidence of light; *monocrystal 2:* V- normal incidence of light; VI- $60^0$ incidence of light (Fig. 6, spectrum III). For monocrystal 1 the ratio of *vibronic* band average intensity to that of the pure electronic band is ~1.0:1.7, indicating strong presence of $\boldsymbol{b_2}$ vibronic bands and consequently of $\boldsymbol{B_2}$ .electronic transitions. For monocrystal 2 this ratio is ~1.0:1.4 (Fig. 6, spectrum V), indicating an even larger projection of the $\boldsymbol{B_2}$ symmetry transition dipole on the developed plane. This observation led us to try aligning the c-crystallographic direction of the sample as close as possible with the direction of the electric field vector. Rotation of monocrystals 1 and 2 by $60^0$ around the axis perpendicular to the electric field vector almost entirely eliminated $\boldsymbol{A_1}$ transitions, revealing predominantly $\boldsymbol{B_2}$ transitions (Fig. 6, panels IV and VI).

The latter transitions, including the lowest energy one of this symmetry (which is not the onset of absorption), form an almost uniform intensity distribution, so there is no resemblance between the molecular and crystal spectra for this type of sample.

## ***Concluding remarks***

This is the first work in which electronic *transmittance* (not reflectance) spectra of π-conjugated organic crystals are obtained in all three crystallographic directions. Spectra of four polymorphs of fluorene and two polymorphs of dibenzofuran were measured in the energy range from the onset of absorption to the ionization energy (nine or ten electronic transitions for each polymorph).

Transitions of $\boldsymbol{A_1}$ ($B_{2u}$ and $B_{3u}$) symmetry (transition dipoles are at oblique angles to the crystallographic axes) in most cases can be traced to corresponding molecular spectra in the whole energy range investigated, irrespective of crystallization conditions.

All lowest energy electronic transitions (at the onset of absorption) of both symmetries in all polymorphs of fluorene and dibenzofuran retain molecular character irrespective of crystallization conditions or the mutual orientation of electronic transition dipoles in the crystal.

The character $\boldsymbol{B_2}$ symmetry transitions above the lowest energy exciton in fluorene and dibenzofuran crystals depend on the crystallization conditions.

In thin crystals, grown from melt between LiF plates under strong uniaxial pressure, the $\boldsymbol{B_2}$ ($B_{1u}$) symmetry electronic transitions retain molecular character. We propose that in this case the crystalline structure of samples is influenced by the LiF substrate, so that predominantly *surface* states contribute to the spectrum.

For polymorph I of dibenzofuran (for which $\boldsymbol{B_2}$ transitions do have molecular character), the lowest-energy transition of $\boldsymbol{B_2}$ ($B_{1u}$) symmetry exhibit hypochromism (and hyperchromism in higher energy transitions) in qualitative agreement with hypochromy/ hyperchromy theory [1-7, 25].

In sublimation flakes and crystal grown from melt under fast crystallization conditions, $\boldsymbol{B_2}$ ($B_{1u}$) symmetry transitions (with parallel transition dipoles in the crystal) fuse together to form an essentially uniform intensity distribution except when the $\boldsymbol{B_2}$ ($B_{1u}$) transition is the onset of absorption. We assume that in these samples substrates do not substantially affect the crystalline structure of the sample; therefore predominantly *bulk* electronic properties are reflected in the spectra of the $\boldsymbol{B_2}$ ($B_{1u}$) transitions of these samples. We do not know of a theoretical model that explains continuous uniform intensity distribution in spectra of multiple transitions in π-conjugated organic crystals.

***Acknowledgments***

The spectra were measured at the National Synchrotron Light Source which is supported by the Office of Basic Energy of the US Department of Energy. We wish to thank Mr. John Trunk for expert technical assistance, Dr. Nykypanchuk for permitting and mentoring the use the polarizing microscope and the atomic force microscope at the Center for Functional Nanomaterials at BNL and for helpful discussions , Ms. Tiffany Bowman for expertl graphical work, Mr. Mike Sullivan for kind guidance and help with use of equipment in the sample preparation room, and Dr. Jacob Padgug for help with preparing the manuscript for publication.

***References***

[1] Tinoco, I . (1960*). J. Chem. Phys*., **33**, 1332; J. Am. Chem. Soc, 82, 4785.

[2] Rhodes, W. (1961*) J. Am. Chem., Soc.*, **83**, 3609.

[3] DeVoe, H. (1964*). J. Chem., Phys.,* **41**, 393

[4] McLachlan, A. D. & Ball, M.A. (1964). *Mol. Phys.*, **8**, 581.

[5] Rhodes, W. & Chase, M. (1967). *Rev. Mod. Phys.*, **39**, 348.

[6] Agranovich, V.M. (2008). *Excitations* in Organic Solids. Oxford Science publishers

[7] Hoffmann, R. (1963*). Radiation Res.,* **20**, 140.

[8] Schellman, J. A. & Schellman, C. (1964). In: *The Proteins*, Neurath H. (Ed.), Academic Press: New York, Vol. 2, 3.

[9] Chandross, E. A., Ferguson, J. & McRae, E. G. (1966). *J. Chem. Phys.,* **45**, 3546.

[10] Mishina, L.A., Sviridova, K. A. Nakhimovsky, L. A. (1975). *Bull. Acad. Sci. USSR*, Phys. Ser., **39**, 134.

[11] Okamoto, K., Itaya, A. & Kusabayashi, S. (1974). *Chemical letters*, Published by Chemical Society of Japan.

[12] Klopffer, W., Rippen, G. & Kowal, J. (1986). *Macromol. Chem., Macromol Symp. ,* **5**, 187.

[13] Bailly, C., Dassonneville, L., Colson, P., Houssier, C., Fukasava, K., Nishimura, S. & Yoshivary, T. (1999*). Cancer Res.,* **59**, 2853.

[14] Bree, A. & Zwarich, R. (1969). *J. Chem. Phys.,* **51**, 903.

[15] Bree, A. Vilcos, V. & Zwarich, R. *J. Mol. Spectr.* (1973). **48**,135.

[16] Popov, K. R., Smirnov, L. V., Grebneva, V. I. & Nakhimovsky, L. (1974*). Opt. Spectrosc,* **37**, 1074.

[17] Igarashi, N., Tajiri, A. & Hatano, M. (1981). *Bull. Chem. Soc. Jpn., 54*, 1511.

[18] Tanaka, M. (1976). Bull. Chem. Soc. Jpn., **49**, 3382.

[19] Gudipati, M. S., Daverkausen, J., Maus, M. & Hohlneicher, G. (1994). *Chem. Phys.,* **186**, 289.

[20] Nguyen, D. D., Trunk, J., Nakhimovsky, L. & Spanget-Larsen, J. (2010*). J.Mol. Spectr.,* **264**, 19-25.

[21] Venghaus, H. & Hinz, H. J. (1974*). Phys.Stat. Sol. (b)*, **65**, 239.

[22] Lahiri, B. N. (1986). Z.Krist., **127**, 456.

[23] Robinson, P. M. & Scott, H. G.(1969). *Mol. Cryst. Liquid. Cryst.,* **5**, 405.

[24] Kurahashi, M., Fukuyo, M., Shimada, A., Furusaki, A. & Nitta, I. (1969). *Bull. Chem. Soc. Japan,* **42**, 2174.

[25] Nakhimovsky, L. A. and Fuchs R. (2007) *Mol. Cryst. Liquid. Cryst.,* **473**, 87. ***Corrigendum (2009), 515, 255***

[26] Grebneva, V. L., Popov, K. R., Smirnov, L. V., Nurmukhametov R. N., & Nakhimovsky, L. (1972*). Opt. Spectrosc,* **33**, 55.

[27] Davidov A.S. Theory of molecular excitons, Plenum Press, New York, 1971

[28] Chakravorty, S. C.; Ganguly, S. C. (1969) *Journal of Physics, B., Proceedings of the Physical Society, Atomic and Molecular Physics*, **2**, 1235-9

[29] Heimel, G; Hammer, K; Ambrosh-Draxl, C; Chunwchirasiri, W; Winocur, M.J.; Hanfland, M; Oerzelt, M. *Physical Review B,* **73**, id. 02410